# Classification of Hyperspectral Images by Using Spectral Data and Fully Connected Neural Network

Z. Dokur, and T. Ölmez

*Abstract*— It is observed that high classification performance is achieved for one- and two-dimensional signals by using deep learning methods. In this context, most researchers have tried to classify hyperspectral images by using deep learning methods and classification success over 90% has been achieved for these images. Deep neural networks (DNN) actually consist of two parts: i) Convolutional neural network (CNN) and ii) fully connected neural network (FCNN). While CNN determines the features, FCNN is used in classification. In classification of the hyperspectral images, it is observed that almost all of the researchers used 2D or 3D convolution filters on the spatial data beside spectral data (features). It is convenient to use convolution filters on images or time signals. In hyperspectral images, each pixel is represented by a signature vector which consists of individual features that are independent of each other. Since the order of the features in the vector can be changed, it doesn't make sense to use convolution filters on these features as on time signals. At the same time, since the hyperspectral images do not have a textural structure, there is no need to use spatial data besides spectral data. In this study, hyperspectral images of Indian pines, Salinas, Pavia centre, Pavia university and Botswana are classified by using only fully connected neural network and the spectral data with one dimensional. An average accuracy of 97.5% is achieved for the test sets of all hyperspectral images.

## I. Introduction

IN hyperspectral images, each pixel contains spectral information obtained in different bands. In this case, a set of spectral images of the investigated geographic area is obtained. In the volumetric image set, the z-axis contains the spectral images. Since the number of spectral bands is more than 100, the number of spectral images in this set is quite large. Therefore, this volumetric image sets are usually large in size (X, Y and Z axis) and also little ground truth data is available for these sets. Researchers use a few ground truth data to label the unknown parts of the region. The large size of the image set increases the computational load and makes it difficult for computers to process these images. Another difficulty in processing these images is that the number of data belonging to the classes is very different. While the number of data for some classes is in the order of 10, the number of data for some classes approaches 10000. This situation adversely affects the training of the networks. Since there is a small number of ground truth data, the number of data belonging to the background is very large and, therefore background data is not used in the classification process. Moreover, since hyperspectral images are acquired from the sky, the effect of water vapor as noise is observed in some spectral bands.

It is observed that taking into account the above difficulties, researchers have proposed many different methods to improve the classification performance of hyperspectral images. The high performance of deep learning methods in recent years has encouraged researchers to use deep learning methods to classify these images [1]. It is observed that hyperspectral images are classified with a success rate of over 95% by using deep learning methods [2–11]. In this context, some researchers have developed complex methods to increase the classification performance by using deep learning methods. [12–18].

In order to increase the classification performance of hyperspectral images, it is observed that the researchers removed some components from the feature vector as they did not contain meaningful information [6, 11 19–23], used spatial information beside the features [24–29], balanced the number of vectors belonging to classes in the training set [2, 6, 8, 14, 16, 18, 29], and tried to use the transfer learning methods [9, 28]. In [6, 11 19–23] studies, the focus has been on improving classification performance by arranging or removing features. In [24–29] studies, spatial data beside spectral data was used and optimum sizes of two- and three-dimensional convolution filters were investigated. This approach naturally increased the computational load. It is observed that the networks are often not properly trained because the number of samples belonging to the classes is very different. This situation also reduces the classification performance. Therefore, the training set is created from the same and few number of samples from each class [2, 6, 8, 14, 16, 18, 29]. In [9, 28] studies, the advantage of transfer learning is used. Because there are so many nodes in these networks, the computational load is very large.

Deep neural networks (DNN) actually consist of two parts: i) Convolutional neural network (CNN) and ii) fully connected neural network (FCNN). While CNN determines the features, FCNN is used in classification. In our previous work [30], it has been proposed to train these two parts separately. For different classification problems, we used the DivFE as a novel structure, and the highest success rates according to the literature were obtained with the minimum number of weights. DivFE only determines features. After DivFE, minimum distance network (MDN) is used instead of fully connected neural network as classifier. The DivFE is used to classify 1D signature vectors of the hyperspectral images, but an average of 92% success was observed. Although DivFE gave high performance with a minimum number of weights for different classification problems, high performance could not be achieved by using the DivFE for hyperspectral images. At this point, the approaches increasing the classification performance began to be investigated.

It is observed that spatial information besides features was used. In hyperspectral images, a single pixel is represented by spectral information acquired from different bands. This spectral information is written in vector form (signature). In this case, there is actually no place to apply 2D convolution filters for single pixel. As a result, the researchers tried to use 1D convolution filters on the feature vector. It is observed that using 1D convolution filters on features did not improve classification performance much [2, 3]. Since these vectors consist of features, the elements of the vector can be reordered among themselves. Reordering is not possible for images (2D) containing spatial information and for time-domain signals (1D). Actually, it is not wrong to use 1D filters on feature; however, the positions of these features in the vector relative to each other must be reordered optimum before training. Researchers then preferred to use spatial information beside spectral features to avoid this confusion [24-29]. Thus, convolution filters are used on the spatial axes (X and Y) instead of spectral features. Spatial information is generally used in textural structures. At the same time, it is very difficult to represent textural characteristics in a hyperspectral image at poor resolution. Despite these problems, this approach has increased the classification performance according to the results obtained by using a 1D-CNN. It is observed that while using 2D convolution filters instead of 1D convolution methods has led to a significant increase in classification performance, using 3D convolution filters instead of 2D convolution methods did not cause a significant increase in classification performance [2, 4, 7, 11, 24, 29]. The reason is hidden in the structure of 2D convolution filters. Fig. 1 shows the movement of 2D and 3D convolution filters. 2D convolution filters have a link to all spectral features, and so 2D filters do not move on the spectral features. 3D convolution filters do not have a full connection to the spectral features, and so 3D filters move on the spectral features [17].

The following reasons lead us to use only fully connected neural network and spectral data: (i) It doesn't make sense to use convolution filters on features, (ii) since hyperspectral images do not have a textural structure, there is no need to use spatial data besides spectral data, and (iii) since spectral data is available as features, it is only necessary to determine classifiers. In a recent study [21], it was observed that high performance was achieved by using only an SVM-based classifier instead of CNN. In the [21], the researchers are focused on determining features for pixels in the hyperspectral images. Therefore, we prefer use only fully connected neural network in DNN structure. However, its structure and its training phase introduces a lot of problems, some of which are as follows: (i) FCNN generally fails to converge to optimum solution, (ii) The fully connected layer contains many more nodes than the convolutional layer (high memory need and high time computing), and (iii) it is difficult to find the coarse structure of FCNN. Our proposed approach has been tested in the "Experiment Results" section to classify five different hyperspectral images (Indian pines, Salinas, Pavia center, Pavia university and Botswana). An average accuracy of 97.5% is achieved for the test sets of all hyperspectral images.

The remainder of this paper is organized as follows: Section 2 explains the structure and training of the proposed FCNN, and the preparation of the datasets. Section 3 presents classification performances on the five different hyperspectral images. In Section 4, the methods proposed in the paper are discussed, and finally, our concluding remarks are mentioned.

## II. METHODS

In this study, hyperspectral images were classified by using only 1D spectral data as features and a fully connected neural network as a classifier. Moreover, in order to strengthen the training, the dataset is balanced by creating an equal number of samples from each class.

### A. Spectral Features for the Hyperspectral Images

Signature vector consists of spectral data. In this study, the size of the vector was not reduced using methods, such as principle component analysis (PCA). The use of spatial data on images with textural structures will increase the classification performances. However, the hyperspectral images do not have textural structures at poor resolution. As shown in Fig. 1, 2D convolution filters have full connection to spectral features. The increase in classification performance is thought to be the result of this situation. For these reasons, we decided to use only spectral data as features. It is observed in the literature that PCA is performed to reduce the size of the signature vector. In the classifier systems, divergence measure which is the ratio of between-class distances to the within-class distances in a feature space is employed to determine the best features [31]. However, while the PCA method uses within-class distances, it ignores between-class distances.

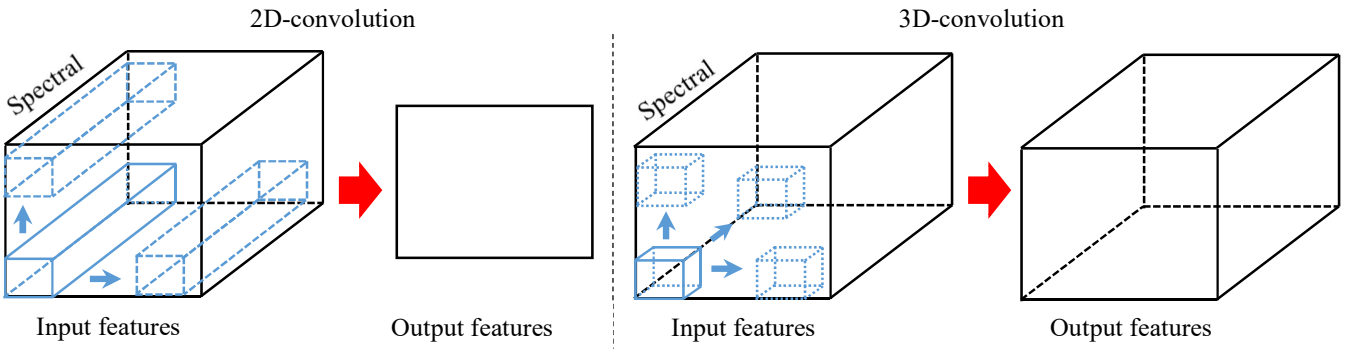

Fig. 1. Illustration of 2-D convolution and 3-D convolution [17].

B. Fully Connected Neural Network as a Classifier

Deep neural networks generally consist of two parts: (i) convolutional neural network representing features and (ii) fully connected neural network as classifier. In this study, the features are already known and consist of spectral data. There is no need to determine features using the convolutional neural network. Therefore, in this study, DNNs classifying hyperspectral images only consist of the FCNN. Fig. 2 shows the structure of the proposed neural network. The signature vector of the hyperspectral image is entered into the input layer of the FCNN. The dimension of the signature vector is different for each hyperspectral image examined. An FCNN with 4 hidden layers was used to classify all hyperspectral images. In this study, the batch normalization and ReLU are used between each layer in FCNN. In the output layer, there are as many nodes as the number of classes. The node with the maximum value represents the class for the input signature vector.

C. Preparation of the Training and the Test Sets

For the networks to be trained properly, there must be an equal number of samples for each class of the training set. Unfortunately, it is observed that there are different numbers of samples for each class of the training set. In this case, it is observed that the researchers try to balance the number of samples for each class of the training set. However, it is extremely important to create a training set for the networks to be properly trained. Unfortunately, detailed information about this process could not be observed in the studies. If the augmentation process is applied to the samples before data sets are separated, the similar samples will exist in both the training and test set. This situation unfairly increases the classification performance. Before augmentation process, the test set must be determined. In this study, 20% of the entire data set is allocated to the test set ($S_{te}$). In this case, an imbalance occurs in the number of samples of the classes in the training set ($S_{tr}$). Among the classes in the $S_{tr}$, the class with the highest sample ($S_{he}$) is found. The number of samples of other classes (excluding $S_{he}$) are increased to the number of samples of $S_{he}$ by randomly duplicating the samples within themselves of classes.

III. EXPERIMENT RESULTS

All simulations were done on a Linux based workstation by using Python. The CPU had 32 cores which are clocked at 2.7 GHz. In addition, the workstation was equipped with a GTX2080 TI graphics card. In this study, hyperspectral images of Indian pines (InP), Salinas (Sal), Pavia centre (PaC), Pavia university (PaU) and Botswana (Bot) are classified by using only the spectral data with one dimensional and the FCNN. In this study, overall accuracy (OA), and average accuracy (AA) are used as success rate. In this section, the tables 2 and 3 show the averages of 30 experiments with randomly formed training and testing sets.

Table 1 shows success rates obtained by using the 1D-CNN, 2D-CNN and 3D-CNN for five different hyperspectral images. It is observed that while using 2D convolution filters instead of 1D convolution methods has led to a significant increase in classification performance, using 3D convolution filters instead of 2D convolution methods did not cause a significant increase in classification performance. In the literature, while 1D-CNN, 2D-CNN and 3D-CNN were compared with their proposed methods, different results have been produced for these classical CNNs [2-4, 11, 19, 26, 27, 29]. This is most likely due to the methods used for training and test set creation, training set balancing and the CNN structure. Moreover, it is observed that 99% classification success is achieved by using very complex methods [7, 10, 11, 14, 27].

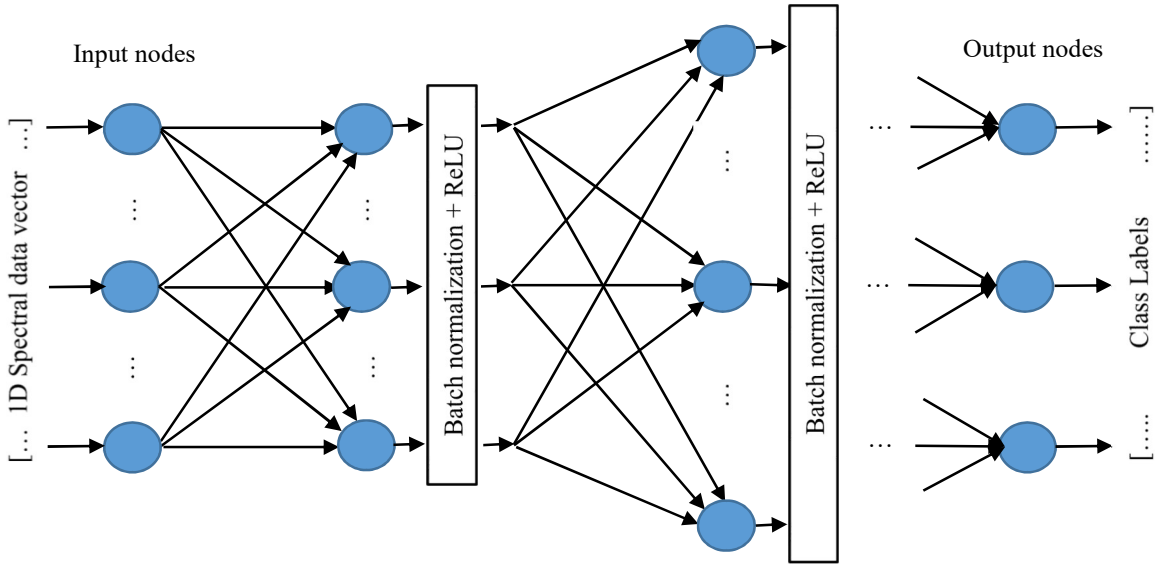

Fig. 2. The structure of the fully connected neural network.

Table 2 shows the success rates (true positive / the number of sample of a class in the test set) for each class of the Indian pines, Salinas and Botswana images by using the proposed FCNN. In the dataset of the Indian pines, the number of samples in classes alfalfa, corn notill, corn mintill, corn, grass pasture, grass trees, grass pasture mowed, hay windrowed, oats, soybean notill, soybean mintill, soybean clean, wheat, woods, building grass trees drives and stone steel towers of the whole dataset is 46, 1428, 830, 237, 483, 730, 28, 478, 20, 972, 2455, 593, 205, 1265, 386 and 93 respectively. 20% of the whole dataset is reserved for the test set. After the balancing process, the number of samples of each class in the training set is 1964 (2455×0.8). The number of nodes in the input layer, first, second, third, fourth hidden layers and output layer are 220, 250, 300, 400, 300 and 16, respectively.

TABLE I
SUCCESS RATES FOR THE HYPERSPECTRAL BY USING THE 1D-CNN, 2D-CNN AND 3D-CNN.

|  | InP % | Sal % | PaC % | PaU % | Bot % | DNN structure |
|---|---|---|---|---|---|---|
| In [2] | OA=90.2 AA=93.3 | OA=91.8 AA=96.7 |  | OA=92.9 AA=93.5 |  | 1D-CNN |
| In [2] | OA=99.1 AA=99.5 | OA=97.5 AA=98.8 |  | OA=99.5 AA=99.4 |  | 2D-CNN |
| In [3] | OA=85.4 AA=79.4 |  |  | OA=93.0 AA=88.6 |  | 1D-CNN |
| In [3] | OA=98.7 AA=98.7 |  |  | OA=99.5 AA=98.8 |  | 2D-CNN |
| In [3] | OA=92.8 AA=89.4 |  |  | OA=95.2 AA=91.2 |  | 3D-CNN |
| In [11] | OA=89.5 AA=86.1 | OA=97.4 AA=98.8 |  | OA=97.9 AA=96.5 |  | 2D-CNN |
| In [11] | OA=91.1 AA=91.6 | OA=93.9 AA=97.0 |  | OA=96.5 AA=97.6 |  | 3D-CNN |
| In [19] | OA=74.8 AA=84.8 | OA=88.6 AA=93.8 |  | OA=82.3 AA=87.3 |  | 2D-CNN |
| In [19] | OA=17.2 AA=27.9 | OA=81.4 AA=87.8 |  | OA=60.8 AA=66.9 |  | 3D-CNN |
| In [29] | OA=86.0 AA=93.3 | OA=88.1 AA=94.6 |  | OA=85.9 AA=85.9 |  | 1D-CNN |
| In [29] | OA=89.1 AA=94.4 | OA=93.6 AA=96.8 |  | OA=91.4 AA=90.5 |  | Mixed 2D-CNN |
| In [29] | OA=82.9 AA=92.3 | OA=91.1 AA=95.7 |  | OA=86.5 AA=85.3 |  | 3D-CNN |
| In our study | OA=93.8 AA=96.0 | OA=95.7 AA=98.3 | OA=99.2 AA=98.3 | OA=96.7 AA=96.4 | OA=98.3 AA=98.6 | 1D-FCNN |

In the dataset of the Salinas, the number of samples in classes broccoli green weeds 1, broccoli green weeds 2, fallow, fallow rough plow, fallow smooth, stubble, celery, grapes untrained, soli vinyard develop, cron senesced gree weeds, lettuce romaine 4wk, lettuce romaine 5wk, lettuce romaine 6wk, lettuce romaine 7wk, vineyard untrained and vineyard vertical trellis of the whole dataset is 2009, 3726, 1976, 1394, 2678, 3959, 3579, 11271, 6203, 3278, 1068, 1927, 916, 1070, 7268 and 1807 respectively. 20% of the whole dataset is reserved for the test set. After the balancing process, the number of samples of each class in the training set is 9016 (11271×0.8). The number of nodes in the input layer, first, second, third, fourth hidden layers and output layer are 224, 250, 300, 400, 200 and 16, respectively.

In the dataset of the Botswana, the number of samples in classes water, hippo grass, floodplain grasses1, floodplain grasses2, reeds1, riparian, firescar2, island interior, acacia woodlands, acacia shrublands, acacia grasslands, short mopane, mixed mopane and exposed soils is 270, 101, 251, 215, 269, 269, 259, 203, 314, 248, 305, 181, 268 and 95 respectively. 20% of the whole dataset is reserved for the test set. After the balancing process, the number of samples of each class in the training set is 251 (314×0.8). The number of nodes in the input layer, first, second, third, fourth hidden layers and output layer are 145, 250, 300, 400, 64 and 14, respectively.

TABLE 2
THE SUCCESS RATES FOR THE HYPERSPECTRAL IMAGES BY USING THE FCNN.

| Indian pines | | Salinas | | Botswana | |
|---|---|---|---|---|---|
| alfalfa | 10/10 | broccoli green weeds 1 | 401/402 | water | 54/54 |
| corn notill | 266/286 | broccoli green weeds 2 | 746/746 | hippo grass | 21/21 |
| corn mintill | 142/166 | fallow | 396/396 | floodplain grasses1 | 51/51 |
| Corn | 46/48 | celery | 716/716 | floodplain grasses2 | 43/43 |
| grass pasture | 95/97 | fallow smooth | 534/536 | reeds1 | 48/54 |
| grass trees | 144/146 | stubble | 790/792 | riparian | 52/54 |
| grass pasture mowed | 6/6 | fallow rough plow | 277/279 | firescar2 | 52/52 |
| hay windrowed | 96/96 | grapes untrained | 1967/2255 | island interior | 41/41 |
| oats | 4/4 | soli vinyard develop | 1241/1241 | acacia woodlands | 62/63 |
| soybean notill | 185/195 | cron senesced gree weeds | 651/656 | acacia shrublands | 49/50 |
| soybean mintill | 446/491 | lettuce romaine 4wk, | 214/214 | acacia grasslands | 60/61 |
| soybean clean | 117/119 | vineyard vertical trellis | 361/362 | short mopane | 37/37 |
| wheat | 40/41 | lettuce romaine 6wk | 184/184 | mixed mopane | 54/54 |
| woods | 244/253 | lettuce romaine 7wk | 212/214 | exposed soils | 19/19 |
| building grass trees drives | 68/78 | vineyard untrained al trellis | 1290/1454 | OA AA | 98.3% 98.6% |
| stone steel towers | 19/19 | lettuce romaine 5wk | 386/386 | | |
| OA AA | 93.8% 96.0% | OA AA | 95.7% 98.3% | | |

Table 3 shows the success rates (true positive / the number of sample of a class in the test set) for each class of the Pavia centre and Pavia university. In the dataset of the Pavia centre, the number of samples in classes water, trees, asphalt, self-blocking bricks, bitumen, tiles, shadows, meadows and bare soil in the whole dataset is 65971, 7598, 3090 2685, 6584, 9248, 7287, 42826 and 2863 respectively. 20% of the whole dataset is reserved for the test set. After the balancing process, the number of samples of each class in the training set is 52778 (65971×0.8). The number of nodes in the input layer, first, second, third, fourth hidden layers and output layer are 102, 250, 300, 400, 200 and 9, respectively.

In the dataset of the Pavia university, the number of samples in classes asphalt, meadows, gravel, trees, painted metal sheets, bare soil, bitumen, self-blocking bricks and shadows in the whole dataset is 6631, 18649, 2099, 3064, 1345, 5029, 1330, 3682 and 947 respectively. 20% of the whole dataset is reserved for the test

set. After the balancing process, the number of samples of each class in the training set is 14919 (18649×0.8). The number of nodes in the input layer, first, second, third, fourth hidden layers and output layer are 103, 250, 300, 400, 200 and 9, respectively.

TABLE 3
THE SUCCESS RATES FOR THE HYPERSPECTRAL IMAGES BY USING THE FCNN.

| Pavia centre | | Pavia University | |
|---|---|---|---|
| water | 13194/13195 | asphalt | 1274/1327 |
| trees | 1499/1520 | meadows | 3679/3730 |
| asphalt | 602/618 | gravel | 387/420 |
| self blocking bricks | 524/537 | trees | 605/613 |
| bitumen | 1281/1317 | painted metal sheets | 269/269 |
| tiles | 1819/1850 | bare soil | 939/1006 |
| shadows | 1402/1458 | bitumen | 260/266 |
| meadows | 8512/8566 | self blocking bricks | 670/737 |
| bare soil | 573/573 | shadows | 190/190 |
| OA | 99.2% | OA | 96.7% |
| AA | 98.3% | AA | 96.4% |

V. Conclusion

In this study, an average accuracy of 97.5% is achieved for five different hyperspectral images. Similar success rates are obtained with the results obtained by using 2D convolution neural networks. In fact, preliminary results validate the accuracy of our hypotheses. In order to increase the classification performances obtained with the FCNN, the coarse structure of the network should be well estimated. Unfortunately, the coarse structure cannot be determined easily because FCNN contains many neurons. It should be noted that a similar structure is used for five different hyperspectral images. In order to increase the classification performance, an optimum structure is not investigated for each different hyperspectral image, and the number of the spectral features are not reduced to determine optimal features.

Because the pixels represent areas of meters in length, the resolution is poor in the hyperspectral images. It is very difficult to represent textural characteristics in a hyperspectral image at this resolution. Actually, the hyperspectral images do not also contain textural features. On the other hand, 2D convolution filters have a link to all spectral features, and so 2D filters do not move on these features. The input of the FCNN have a link to all spectral features, as in 2D-CNN. For this reason, high classification performance has been achieved with 2D-CNNs. Moreover, it is observed that the classification performances obtained with 3D-CNN are lower than those obtained with 2D-CNNs. In both 2D-CNN and 3D-CNNs, the computational load increases because their structures grow.

It is extremely important to make a pre-processing on the data in order to increase the success rate of the hyperspectral images. Generally, two types of studies are observed in the literature: 1) balancing the number of samples of the classes in the training set and 2) the use of the dimension reduction methods (such as PCA) to eliminate unnecessary features. In this study, the number of samples of classes in the training set is balanced to improve the training of the network. This approach resulted in an increase of approximately 6 or 7 points on the success rates. As stated in the method section, after the training and test sets are separated, the number of samples of the classes in the training set should be balanced. When all classes are balanced before separating training and test sets, we achieved average accuracy of 100%. It is observed that data reduction methods, such as PCA, are frequently used. while the PCA method uses within-class distances, it ignores between-class distances. In the classification problems, the divergence value computed over a feature set, points to the effectiveness of the features in terms of being intra-class representative and inter-class discriminative [31]. A high divergence value signifies that a favorable distribution of the feature vectors has been attained. The size of the signature vectors is reduced from 102 to 30 by maximizing the divergence value for the Pavia center image. In this case, the same performance is achieved for the Pavia center image. In FCNN structure, dimension reduction affects only the input layer, whereas in CNN structure, every layer is affected by dimension reduction process. In the next study, we will determine the best features using methods that maximize the divergence value [31].

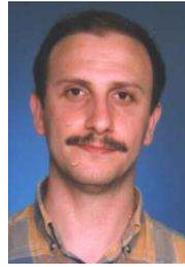

**Tamer Ölmez** received the B.Sc. degree in electrical and electronics engineering in 1985, the M.Sc. degree in computer engineering in 1988, and the Ph.D. degree in biomedical engineering in 1995, all from Istanbul Technical University, Turkey. Between 1985-1988, he worked as a research engineer at Alcatel Turkey. Until the end of 1989 he worked at The Scientific and Technical Research Council of Turkey as a research engineer. Since 1991 he has been with the Department of Electrical and Electronics Engineering at Istanbul Technical University, Turkey, where he is a professor since 2003. His current research interests are machine learning (deep neural networks, genetic algorithms, fuzzy logic), pattern recognition, biomedical instrumentation, bioinformatics, medical informatics, biological signal and image processing, embedded system design by Linux OS/FPGA/DSP/microcontrollers, android and internet applications by using QT developer, developing application on embedded Linux systems (the RaspberryPi etc.) using C/Python/PyQT/Opencv/Tensorflow.

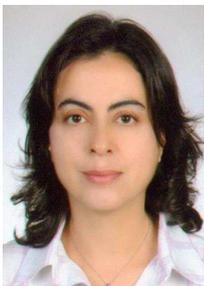

**Zümray Dokur** received the B.Sc., M.Sc., and Ph.D. degrees from Istanbul Technical Uni- versity, Turkey, all in electronics and communication engineering. Since 1992 she has been with the Department of Electronics and Communication Engineering at Istanbul Technical University, Turkey, where she is a professor at present. She serves as an associate editor of Neural Processing Letters since 2008. Her research interests include pattern recognition, biomedical signal and image processing, deep neural networks, bioinformatics, and brain-computer interface designs using electroencephalogram.